\documentclass[11pt,twoside]{article}

\usepackage{asp2010}
\usepackage{epsf}
\usepackage{graphicx} 
\usepackage{lscape}

\begin{document}
\title{Asymmetric Structure of Quiescent Filament Channels Observed by Hinode/XRT and STEREO/EUVI}   
\author{Yingna Su, Adriaan Van Ballegooijen, and Leon Golub}   
\affil{Harvard-Smithsonian Center for Astrophysics}    

\begin{abstract}  

We present a study on the structure of quiescent filament channels observed by Hinode/XRT and STEREO/EUVI from December  2006 to February 2009.  For 10 channels identified on the solar disk, we find that the emission on the two sides of the channel is asymmetric in both X-rays and EUV: one side has curved bright features while the other side has straight faint features. We interpret the results in terms of a magnetic flux rope model. The asymmetry in the emission is due to the variation in axial magnetic flux along the channel, which causes
one polarity to turn into the flux rope, while the field lines from the other polarity are open or 
connected to very distant sources. For 70 channels identified by cavities at the limb, the asymmetry cannot be clearly identified.

\end{abstract}
 
\section{Observations of Quiescent Filament Channels}
 
Filament channels are defined as regions in the chromosphere surrounding a Polarity Inversion Line (PIL) where the chromospheric fibrils are aligned with the PIL, indicating the presence of an axial magnetic field  \citep{1971SoPh...19...59F, 1971SoPh...20..298F, 1998ASPC..150..257G, 1994ssm..work..303M}. In this paper we study filament channels on the quiet Sun. The studied channels are divided into two types: Type I channels are identified based on observed long and continuous H$\alpha$ filaments; while Type II channels are identified according to the cavities at the limb observed by XRT. 

We first select 10 Type I filament channels during November 2006 and December 2008. These channels are identified based on the H$\alpha$ observations (mainly provided by Kanzelh\"{o}he Solar Observatory or Mauna Loa Solar Observatory) of long and continuous filaments. Type I channels are usually located in low-latitude active region remnants. A list of Type I channels is shown in Table 1. These channels can be divided into 5 groups, and the same channel at different solar rotations is classified into one group. Only five possible corresponding cavities are observed by XRT or TRACE, while the filament channels are at the limb. The visibility of cavity may be affected by the direction of the filament channels or some bright active regions close by.

\begin{table}[!ht]
\caption{Type I Filament Channels.}
\smallskip
\begin{center}
{\small
\begin{tabular}{ccc|ccc}
\tableline
\noalign{\smallskip}
Channel &  & Cavity & Channel &   & Cavity\\
     Date/Hemisphere  & Group & Date &Date/Hemisphere & Group & Date \\
\noalign{\smallskip}
\tableline
\noalign{\smallskip}
2006-11-01/S & 1 & 10-27 & 2008-03-02/N+S & 3 & No\\
2006-12-26/S & 1  & 12-19 & 2008-03-27/S & 3 & No\\
2007-01-19/S & 1  &  No & 2008-04-24/S & 3& 05-02\\
2007-04-30/N & 2 & No & 2008-11-29/N &4& 11-24\\
2008-02-02/N+S & 3 & No & 2008-12-09/N & 5 & 12-03\\
\noalign{\smallskip}
\tableline
\end{tabular}
}
\end{center}
\end{table}

\begin{figure}[!ht]
\begin{center}
\includegraphics[scale=0.5]{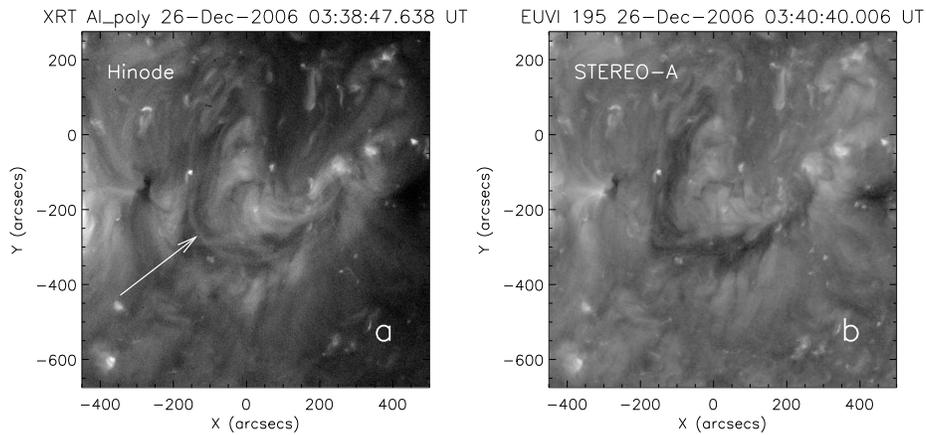}
\end{center}
\caption{Hinode/XRT (a) and STEREO/EUVI (b) images of a filament channel on 2006 Dec 26.}\label{fig1}
\end{figure}

Filament Channels on the quiet Sun are often observed as dark channels in X-rays and EUV, and an example 
is shown in Figure 1. Sometimes, sheared loops (indicated by white arrows in Figure 1a) within the filament channel are observed in X-rays, but not in EUV. Some sheared loops are stable, while others are transient structures. This result is consistent with the findings by \citet{1998ASPC..150..171S} using Yohkoh/SXT observations.

\begin{figure}[!ht]
\begin{center}
\includegraphics[scale=0.5]{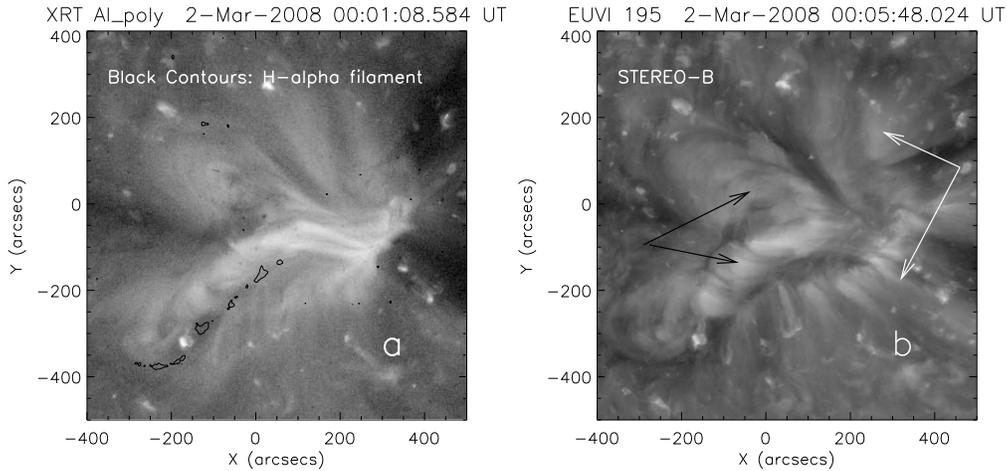}
\end{center}
\caption{Hinode/XRT (a) and STEREO/EUVI (b) images of a filament channel on 2008 March 2.}\label{fig2}
\end{figure}

We find that the structure on the two sides of the channels is asymmetric in both X-rays and EUV. An example of a
filament channel observed on 2008 March 2  is shown in Figure 2. This figure shows that the eastern side (black arrow) has curved bright features, while the western side (white arrows) has straight faint features. This asymmetric structure is found for all of the Type I filament channels.

 \begin{figure}[!ht]
\begin{center}
\includegraphics[scale=0.5]{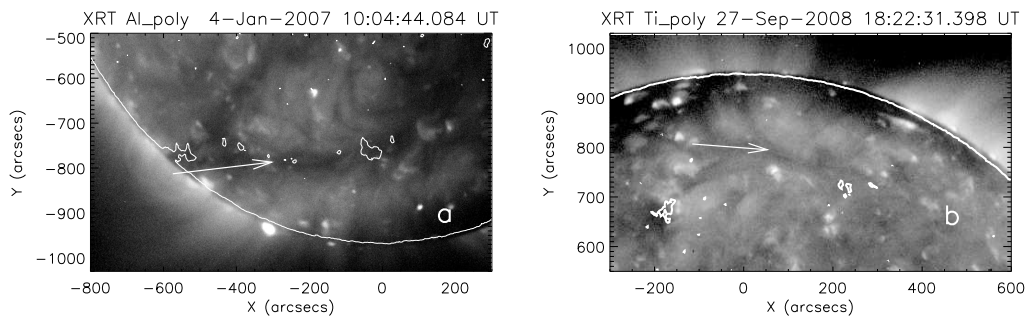}
\end{center}
\caption{Hinode/XRT images of two Type II filament channels. The white contours refer to the corresponding 
H$\alpha$ filament.}\label{fig3}
\end{figure}

We then identify 70 type II filament channels based on cavities in XRT synoptic observations, which are taken twice a day normally. First we look for cavities from these synoptic images.  Then the filament channels are identified as the disk counterpart from the images taken 7 days earlier and later, for the cavities on the west and east limb, respectively.

The H$\alpha$ filaments in these channels are shown as dotted or short segments.  The emission on the two sides of the channels is weaker than that in Type I channels. Based on the structure on the two sides of the channel, the 16 well observed type II channels can be divided into two types: curved on the disk side and straight or unclear on the polar side (Figure 3a); straight or unclear on the disk side and straight on the polar side (Figure 3b). Only one out of the 16 well observed channels appears to show asymmetric structure on the two sides of the channel (see Figure 3a).

\section{Interpretation}
 
 \begin{figure}[!ht]
\begin{center}
\includegraphics[scale=0.5]{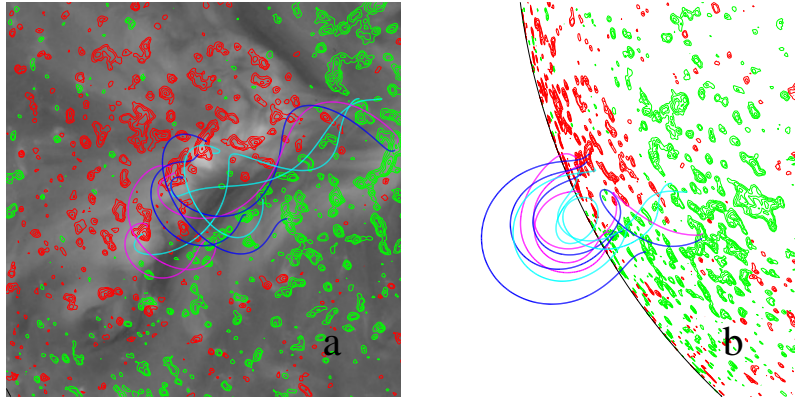}
\end{center}
\caption{Front (a) and side (b) views of selected field lines of a NLFFF model for the filament channel on
2008 Mar 1. The black and white contours refer to the positive and negative polarities observed by SOLIS. }\label{fig4}
\end{figure}

The asymmetric structure of the Type I channels is likely due to the fact that these channels are inclined with respect to the lines of constant latitude, and the fields are stronger at the equatorial end of the channel. We suggest that the channel contains a weakly twisted flux rope, and the axial flux of the flux rope increases towards the equator, so that new flux is drawn into the flux rope from one side of the PIL. The bright curved features correspond to the lower parts of the field lines that turn into the flux rope, while the straight faint features are the lower parts of the overlying coronal arcade. The Type II channels are usually East-West oriented, and the axial flux may be relatively constant along such channels, so asymmetry is rarely observed.

We construct a nonlinear force-free field (NLFFF) model of a Type I channel using the flux rope insertion method \citep{2004ApJ...612..519V}. Several flux strands with different fluxes and lengths along the PIL are inserted into a potential field model, and magneto-frictional relaxation is applied to produce a NLFFF model for the channel on 2008 March 1. Front (a) and side (b) views of selected field lines are shown in Figure 4. Note that field lines are drawn into the flux rope from the North-East side where the bright curved EUV features are located.

In summary, we have observed the fine structure of filament channels on the quiet Sun, and we find that for Type I channels the EUV and X-ray emission is asymmetric on the two sides of the channel. The results can be interpreted in terms of a flux rope model with the magnitude of the axial flux increasing along the channel as we approach the equator. The same effect is rarely found for East-West oriented polar crown channels.
 
Acknowledgements: The authors wish to thank the team of Hinode/XRT, STEREO, TRACE, SOLIS, MLSO, KSO for providing the valuable data. US members of the XRT team are supported by NASA contract NNM07AB07C to Smithsonian Astrophysical Observatory (SAO). \textrm{Hinode} is a Japanese mission developed and launched by ISAS/JAXA, with NAOJ as domestic partner and NASA and STFC (UK) as international partners. It is operated by these agencies in co-operation with ESA and the NSC (Norway). 


\end{document}